\documentclass[conference]{IEEEtran}
\usepackage{etex}
\usepackage{ifpdf}
\ifpdf
\usepackage[draft,pdfborder={0 0 0}]{hyperref}
\fi

\usepackage{cite}
\usepackage{graphicx}

\usepackage[cmex10]{amsmath}
\usepackage{nicefrac}

\usepackage{bm}
\usepackage[varg]{txfonts}
\let\mathbb=\varmathbb
\DeclareSymbolFont{letters}{OML}{ztmcm}{m}{it}

\usepackage{array}
\usepackage[caption=false,font=footnotesize]{subfig}
\usepackage{tabularx}
\usepackage{multirow}
\usepackage{dcolumn}
\usepackage{booktabs}

\usepackage{fixltx2e}
\usepackage{stfloats}

\usepackage{color}
\newcommand{\fixme}[2]{\ifx&#2&{\color{red}#1}\else{\color{red}FIXME\{}#1{\color{red}\}}\footnote{{\color{red}#2}}\PackageWarning{Fixme}{#1: #2}\fi}

\usepackage{tikz,pgf}
\usepackage{pgfplots}
\usetikzlibrary{shapes,positioning,arrows,decorations.markings,fit,calc,patterns}

\title{Hardware Decoders for Polar Codes: An Overview}

\author{\IEEEauthorblockN{
Pascal Giard\IEEEauthorrefmark{1}$^\diamond$, %
Gabi Sarkis\IEEEauthorrefmark{1}, %
Alexios Balatsoukas-Stimming\IEEEauthorrefmark{2}, %
YouZhe Fan\IEEEauthorrefmark{3},\\%
Chi-ying Tsui\IEEEauthorrefmark{3}, %
Andreas Burg\IEEEauthorrefmark{2}, %
Claude Thibeault\IEEEauthorrefmark{4}, and %
Warren J. Gross\IEEEauthorrefmark{1}}\vspace{2pt}
  \IEEEauthorblockA{\IEEEauthorrefmark{1}Dept. of Electr. and Comput. Eng., McGill Univ., Montr\'eal, Qu\'ebec, Canada.}%
  \IEEEauthorblockA{\IEEEauthorrefmark{2}Telecommun. Circuits Lab., EPFL, Lausanne, Switzerland.}%
  \IEEEauthorblockA{\IEEEauthorrefmark{3}Dept. of Electron. and Comput. Eng., Hong Kong Univ. of Sci. and Technol., Hong Kong, China.}%
  \IEEEauthorblockA{\IEEEauthorrefmark{4}Dept. of Electr. Eng., \'Ecole de technologie sup\'erieure, Montr\'eal, Qu\'ebec, Canada.}%
\IEEEauthorblockA{$^\diamond$Email: pascal.giard@mail.mcgill.ca}}

\begin{document}

\maketitle

\begin{abstract}
Polar codes are an exciting new class of error correcting codes that achieve the symmetric capacity of memoryless channels. Many decoding algorithms were developed and implemented, addressing various application requirements: from error-correction performance rivaling that of LDPC codes to very high throughput or low-complexity decoders. In this work, we review the state of the art in polar decoders implementing the successive-cancellation, belief propagation, and list decoding algorithms, illustrating their advantages.
\end{abstract}

\section{Introduction}
\label{sec:intro}
Polar codes are the first codes with an explicit construction to asymptotically achieve the symmetric capacity of memoryless channels using a low-complexity, successive-cancellation (SC), decoding algorithm\cite{Arikan2009}. Additionally, they were shown to be free of error floors when used with binary-input symmetric memoryless channel, and are therefore attractive for wired communications and storage systems \cite{Mondelli2015}.

The serial nature of SC decoding limits the throughput of its implementations. Two approaches are used in literature to overcome the sequential nature of SC: exploiting the polar code structure to estimate multiple bits in parallel while still using SC-based algorithms \cite{Alamdar-Yazdi2011,Sarkis2014}, and using the belief propagation (BP) decoding algorithm with parallel message passing \cite{Hussami2009,Park2014}. Hardware decoders implementing these algorithms reach throughputs of multiple Gbps and can exceed 100 Gbps when unrolling is used \cite{Giard_IET_2015,Giard_JSAC_2015}.

While SC and BP based decoders are fast and efficient, their error-correction performance can be inferior to that of other modern codes such as low-density parity-check (LDPC) codes. However, when polar codes are decoded using the successive cancellation list (SCL) decoding algorithm~\cite{Tal2015}, their error-correction performance was shown to exceed that of LDPC codes used in recent wireless communication standards \cite{Tal2015}.

In this paper, we review the state of the art in hardware polar decoder implementations. We start with a review of polar codes in Section~\ref{sec:background}. SC-based implementations are discussed in Section~\ref{sec:sc-based} and BP-based decoders in Section~\ref{sec:bp}. Finally, SCL decoder implementations are reviewed in Section~\ref{sec:list}.


\section{Background}
\label{sec:background}
Polar codes are recursively constructed from a $2\times 2$ polarizing transformation $F = \left[\begin{smallmatrix} 0 & 1\\1 & 1\end{smallmatrix}\right]$: a vector $u_0^1$ of two bits $u_0$ and $u_1$ is encoded using $F$ to yield a polar codeword. When estimating $u_0$ and $u_1$ using SC decoding, the probability of correctly estimating $u_0$ decreases while that of $u_1$ increases compared to an uncoded vector. This transform is recursively applied $\log N$ times to encode $N$ bits. As $N \to \infty$ the probability of correct detection approaches 1.0 (reliable) or 0.5 (unreliable) and the proportion of reliable bits approaches the symmetric capacity of the underlying memoryless channel.

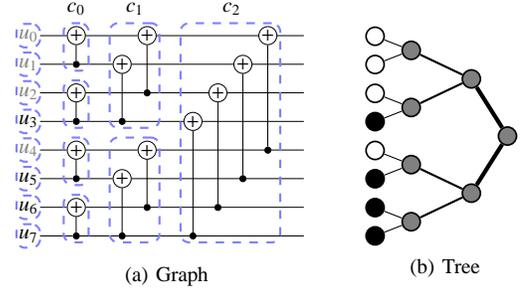
\begin{figure}[t]
  \centering
  \subfloat[Graph]{\label{fig:sc-graph}\scalebox{0.8}{\newcommand{\ubit}[1]{$u_{#1}$}
\newcommand{\fbit}[1]{\color{gray}$u_{#1}$}
\begin{tikzpicture}[baseline=(s37.center)]

\usetikzlibrary{shapes,positioning,arrows,decorations.markings,fit}

\definecolor{varnode_fill}{RGB}{0,0,0}
\definecolor{chknode_fill}{RGB}{255,255,255}

\tikzset{
  chknode/.style={draw,fill=chknode_fill,circle,minimum size=0.3cm, inner sep=0},
  varnode/.style={draw,fill=varnode_fill,circle,minimum size=0.1cm, inner sep=0},
  sep/.style={rectangle,minimum width=0.25cm, inner sep=0},
  bit/.style={circle, inner sep = 0}
}

\tikzset{blue dotted/.style={draw=blue!50!white, line width=1pt,
    dash pattern=on 4pt off 4pt,
    inner sep=0.5mm, rectangle, rounded corners}};

\tikzset{blue dotted tight/.style={draw=blue!50!white, line width=1pt,
    dash pattern=on 4pt off 4pt,
    inner sep=0mm, rectangle, rounded corners}};

\matrix[row sep=1mm, column sep=1mm] {
	\node[bit] (n0s0) {\fbit{0}}; & \node[sep] (s0) {}; & \node[chknode] (n0s1) {$+$}; & \node[sep] (s10) {}; &&	\node[chknode] (n0s2) {$+$}; &	\node[sep] (s20) {}; &&&& \node[chknode] (n0s3) {$+$}; &    \node[sep] (s30) {}; & \node[circle] (n0s4) {}; \\ 
	\node[bit] (n1s0) {\fbit{1}}; & \node[sep] (s1) {}; & \node[varnode] (n1s1) {};	   & \node[sep] (s11) {}; &	 \node[chknode] (n1s2) {$+$}; && \node[sep] (s21) {}; &&&  \node[chknode] (n1s3) {$+$}; &&   \node[sep] (s31) {}; & \node[circle] (n1s4) {}; \\ 
	\node[bit] (n2s0) {\fbit{2}}; & \node[sep] (s2) {}; & \node[chknode] (n2s1) {$+$}; & \node[sep] (s12) {}; &&	 \node[varnode] (n2s2) {};    &	 \node[sep] (s22) {}; &&   \node[chknode] (n2s3) {$+$}; &&&  \node[sep] (s32) {}; & \node[circle] (n2s4) {}; \\ 
	\node[bit] (n3s0) {\ubit{3}}; & \node[sep] (s3) {}; & \node[varnode] (n3s1) {};	   & \node[sep] (s13) {}; &	 \node[varnode] (n3s2) {};    && \node[sep] (s23) {}; &	   \node[chknode] (n3s3) {$+$}; &&&& \node[sep] (s33) {}; & \node[circle] (n3s4) {}; \\ 
	\node[bit] (n4s0) {\fbit{4}}; & \node[sep] (s4) {}; & \node[chknode] (n4s1) {$+$}; & \node[sep] (s14) {}; &&	 \node[chknode] (n4s2) {$+$}; &	 \node[sep] (s24) {}; &&&& \node[varnode] (n4s3) {};	&    \node[sep] (s34) {}; & \node[circle] (n4s4) {}; \\ 
	\node[bit] (n5s0) {\ubit{5}}; & \node[sep] (s5) {}; & \node[varnode] (n5s1) {};	   & \node[sep] (s15) {}; &	 \node[chknode] (n5s2) {$+$}; && \node[sep] (s25) {}; &&&  \node[varnode] (n5s3) {};	&&   \node[sep] (s35) {}; & \node[circle] (n5s4) {}; \\ 
	\node[bit] (n6s0) {\ubit{6}}; & \node[sep] (s6) {}; & \node[chknode] (n6s1) {$+$}; & \node[sep] (s16) {}; &&	 \node[varnode] (n6s2) {};    &	 \node[sep] (s26) {}; &&   \node[varnode] (n6s3) {};	&&&  \node[sep] (s36) {}; & \node[circle] (n6s4) {}; \\ 
	\node[bit] (n7s0) {\ubit{7}}; & \node[sep] (s7) {}; & \node[varnode] (n7s1) {};	   & \node[sep] (s17) {}; &	 \node[varnode] (n7s2) {};    && \node[sep] (s27) {}; &	   \node[varnode] (n7s3) {};	&&&& \node[sep] (s37) {}; & \node[circle] (n7s4) {}; \\ 
};
\path[-] (n0s0) edge (n0s1) (n0s1) edge (n0s2) (n0s2) edge (n0s3) (n0s3) edge (n0s4);
\path[-] (n1s0) edge (n1s1) (n1s1) edge (n1s2) (n1s2) edge (n1s3) (n1s3) edge (n1s4);
\path[-] (n2s0) edge (n2s1) (n2s1) edge (n2s2) (n2s2) edge (n2s3) (n2s3) edge (n2s4);
\path[-] (n3s0) edge (n3s1) (n3s1) edge (n3s2) (n3s2) edge (n3s3) (n3s3) edge (n3s4);
\path[-] (n4s0) edge (n4s1) (n4s1) edge (n4s2) (n4s2) edge (n4s3) (n4s3) edge (n4s4);
\path[-] (n5s0) edge (n5s1) (n5s1) edge (n5s2) (n5s2) edge (n5s3) (n5s3) edge (n5s4);
\path[-] (n6s0) edge (n6s1) (n6s1) edge (n6s2) (n6s2) edge (n6s3) (n6s3) edge (n6s4);
\path[-] (n7s0) edge (n7s1) (n7s1) edge (n7s2) (n7s2) edge (n7s3) (n7s3) edge (n7s4);

\path[-] (n0s1) edge (n1s1);
\path[-] (n2s1) edge (n3s1);
\path[-] (n4s1) edge (n5s1);
\path[-] (n6s1) edge (n7s1);

\path[-] (n0s2) edge (n2s2);
\path[-] (n1s2) edge (n3s2);
\path[-] (n4s2) edge (n6s2);
\path[-] (n5s2) edge (n7s2);

\path[-] (n0s3) edge (n4s3);
\path[-] (n1s3) edge (n5s3);
\path[-] (n2s3) edge (n6s3);
\path[-] (n3s3) edge (n7s3);

\node (g_n0s0) [blue dotted tight, fit = (n0s0)] {};
\node (g_n1s0) [blue dotted tight, fit = (n1s0)] {};
\node (g_n2s0) [blue dotted tight, fit = (n2s0)] {};
\node (g_n3s0) [blue dotted tight, fit = (n3s0)] {};
\node (g_n4s0) [blue dotted tight, fit = (n4s0)] {};
\node (g_n5s0) [blue dotted tight, fit = (n5s0)] {};
\node (g_n6s0) [blue dotted tight, fit = (n6s0)] {};
\node (g_n7s0) [blue dotted tight, fit = (n7s0)] {};

\node (g_n0s1) [blue dotted, fit = (n0s1) (n1s1)] {};
\node (g_n1s1) [blue dotted, fit = (n2s1) (n3s1)] {};
\node (g_n2s1) [blue dotted, fit = (n4s1) (n5s1)] {};
\node (g_n3s1) [blue dotted, fit = (n6s1) (n7s1)] {};

\node (g_n0s2) [blue dotted, fit = (n0s2) (n1s2) (n2s2) (n3s2)] {};
\node (g_n1s2) [blue dotted, fit = (n4s2) (n5s2) (n6s2) (n7s2)] {};

\node (g_n0s3) [blue dotted, fit = (n0s3) (n1s3) (n2s3) (n3s3) (n4s3) (n5s3) (n6s3) (n7s3)] {};

\node at ($(n0s1)+(0,0.45)$) (c0) {$c_0$};
\node at ($(n0s2)+(-0.2,0.45)$) (c1) {$c_1$};
\node at ($(n0s3)+(-0.6,0.45)$) (c2) {$c_2$};

\end{tikzpicture}}}
  \quad
  \subfloat[Tree]{\label{fig:sc-tree}\scalebox{0.8}{\begin{tikzpicture}[baseline = (0_7.center),
        level/.style={level distance = 6mm},
        level 1/.style={sibling distance=19mm, edge from parent/.style={draw,black,line width=2pt}},
        level 2/.style={level distance=10mm, sibling distance=9.5mm, edge from parent/.style={draw,black,line width=1pt}},
        level 3/.style={sibling distance=4.7mm, edge from parent/.style={draw,black,line width=0.5pt}},
        ]

\tikzset{
frozen/.style={thick,draw=black,fill=white,minimum size=3mm,circle, inner sep=0},
fullspace/.style={thick,draw=black,fill=black,minimum size=3mm,circle, inner sep = 0},
mixed/.style={thick,draw=black,fill=gray,minimum size=3mm,circle, inner sep = 0},
ml_mixed/.style={thick,draw=black,fill=blue,minimum size=3mm,circle, inner sep = 0}
}

\node[mixed] (p){} [grow=left]
	child {node[mixed] (2_0){}
		child {node[mixed] (1_0){}
			child {node[frozen] (a0_0){}
			}
			child {node[frozen] (a0_1){}
			}
		}
		child {node[mixed] (1_2){}
			child {node[frozen] (0_2){}
			}
			child {node[fullspace] (0_3){}
			}
		}
	}
	child {node[mixed] (v){}
		child {node[mixed] (cl){}
			child {node[frozen] (0_4){}
			}
			child {node[fullspace] (0_5){}
			}
		}
		child {node[mixed] (cr){}
			child {node[fullspace] (0_6){}
			}
			child {node[fullspace] (0_7){}
			}
		}
	}
;

\end{tikzpicture}}}
  \vspace{-5pt}
  \caption{Graph (a) and tree (b) representation of an (8, 4) polar code.}
  \label{fig:pc8}
\end{figure}

To construct an ($N$, $k$) polar code, the $k$ most reliable bits in $u_0^{N-1}$ are used to carry the information bits; while the remaining bits are frozen by setting them to a predetermined value---usually `0'. Fig.~\ref{fig:sc-graph} shows the graph representation of an (8, 4) polar code where the frozen and information bits are labeled in gray and black, respectively.
Due to the recursive nature of polar code construction, binary trees are a natural representation for these codes. In Fig.~\ref{fig:sc-tree}, the white (black) leaf nodes correspond to frozen (information) bits; whereas the gray nodes correspond to the polar transformations encircled in Fig.~\ref{fig:sc-graph}. Each sub-tree rooted at a node of depth $\log_2N_v$, where leaf nodes have a depth of 0, corresponds to constituent polar codes of length $N_v$.

Quantization in hardware decoders varies based on polar code length and the decoding algorithm used. Many implementations, e.g. \cite{Sarkis2014}, use fewer bits for channel reliability information than for internal values. SCL decoders \cite{Balatsoukas2015} are less tolerant of value saturation and therefore require more quantization bits than their SC counterparts. Finally, longer codes require more integer bits to represent the wider range of their internal values.


\section{Successive Cancellation-Based Decoders}
\label{sec:sc-based}
In this section we briefly go over the algorithms and architectures that led to the fastest hardware decoder implementations based on the successive-cancellation (SC) algorithm.

\subsection{SC-based Decoding Algorithms}
The SC decoding algorithm traverses the entire polar code tree, e.g. Fig.~\ref{fig:sc-tree}, depth first, visiting all leaf nodes.
To reduce latency, the simplified successive-cancellation (SSC) decoding algorithm does not traverse sub-trees whose leaves all correspond to frozen or information bits. Instead, it applies a decision rule immediately \cite{Alamdar-Yazdi2011}. Similarly, constrained maximum-likelihood decoding of multiple bits can be employed to trim the decoder tree \cite{Sarkis2013,Yuan2014,Xiong2015}.

The Fast-SSC decoding algorithm extends the SSC algorithm by applying low-complexity decoding rules when encountering certain types of sub-trees \cite{Sarkis2014,Giard_SIPS_2015}. Specialized decoding of repetition and single-parity-check (SPC) codes are the most notable examples and reduce the decoder-tree size, significantly reducing the number of calculations and increasing the decoding speed.

\subsection{Fast-SSC Decoders}

The configurable hardware implementation of the Fast-SSC algorithm resembles a processor \cite{Sarkis2014}. It features memory for the soft and hard internal values and buffers to allow uninterrupted operation while the next frame is loaded and the previously estimated codeword offloaded. The decoder accepts a set of instructions representing the desired polar code.
These instructions are utilized by the controller to generate the load and store addresses as well as the `select' signals to route the data in and out of the different processing units.

\subsection{Unrolled Decoders}
First applied to polar decoders in \cite{Giard_IET_2015} and \cite{Dizdar2015}, improved and generalized in \cite{Giard_JSAC_2015}, unrolling is a technique that has been successfully applied to other types of decoders before, such as the high-speed LDPC decoders of \cite{Schlafer2013}.

An unrolled polar decoder instantiates processing elements for each and every node in the decoder tree of a specific polar code. This way, each processing element can process a different received vector.
By inserting registers at each decoder stage, a new frame can be loaded and an estimated codeword output at every clock cycle. While this deeply-pipelined architecture provides very high throughput, it requires a significant amount of memory for data persistence that increases with the code length. As a compromise, an initiation interval $\mathcal{I}$ greater than 1 can be defined where a new frame is fed to the decoder every $\mathcal{I}$ clock cycles. The period at which estimated codewords are output is also of $\mathcal{I}$ clock cycles.

\subsection{Implementation Results}
{
\setlength{\tabcolsep}{0.1cm}
\begin{table}
  \centering
  \caption{Results for SC-based polar decoders of various code lengths ($N$) and rates ($R$) implemented on a Altera Stratix IV EP4SGX530KH40C2 FPGA.}
  \vspace{-2.5pt}
  \begin{tabular}{cccrrrccc}
    \toprule
    \multirow{2}{*}{Impl.} & \multirow{2}{*}{$N$} & \multirow{2}{*}{$R$} & \multicolumn{1}{c}{\multirow{2}{*}{LUTs}} & \multicolumn{1}{c}{\multirow{2}{*}{Regs.}} & \multicolumn{1}{c}{RAM} & $f$ & T/P & Latency\\
    & & & & & \multicolumn{1}{c}{\footnotesize(kbits)} & {\footnotesize(MHz)} & {\footnotesize(Gbps)} & {\footnotesize($\mu$s)}\\
    \midrule
  \cite{Sarkis2014}      &32,768& 0.9 & 25,866  &  7,209  &   536 & 108 & \phantom{00}1.2 &26.4\vspace{1.2pt}\\
  \cite{Giard_SIPS_2015} & 1,024 & 0.5 & 24,821  &  5,823  &    36 & 103 & \phantom{00}0.6 & \phantom{0}1.6\vspace{1.2pt}\\
  \cite{Giard_JSAC_2015} & 1,024 & 0.5 &  86,998 &  65,618 &     0 & 218 & \phantom{00}4.5 & \phantom{0}1.7 \\
                         & 1,024 & 0.5 & 136,874 & 188,071 &    84 & 248 & 254.1 & \phantom{0}1.5 \\
                         & 2,048 & 0.5 & 217,175 & 261,112 & 5,362 & 203 & 415.7 & \phantom{0}3.2 \\
    \bottomrule
  \end{tabular}
  \label{tab:impl:sc}
\end{table}
}

Table~\ref{tab:impl:sc} shows FPGA implementation results for both configurable and unrolled Fast-SSC decoders. The decoder of \cite{Sarkis2014} was the first polar decoder to reach a throughput of 1~Gbps. The decoder of \cite{Giard_SIPS_2015} is an improvement over \cite{Sarkis2014} where support for other constituent codes was added to improve the throughput in decoding lower-rate codes. The last three rows of Table~\ref{tab:impl:sc} are results for the unrolled decoders of \cite{Giard_JSAC_2015} showing that throughputs in the hundreds of Gbps are achievable at the cost of area and resource usage.


\section{Belief Propagation Decoders}
\label{sec:bp}
Belief propagation (BP) decoding of polar codes is a message passing algorithm over the graph representation \cite{Arikan2009}, achieving similar error-correction performance to SC decoding. Soft messages are iteratively propagated in the graph until a stopping criterion is met e.g. the maximum number of iterations is reached. Then, threshold detection is applied to the left-hand-side messages to generate the estimated codeword.
This section reviews different design aspects of BP decoders.

\subsection{Fast BP-Based Decoders} 
\label{subsec:LL_BPD}
From Fig.~\ref{fig:sc-graph}, the graph of polar codes of length $N$ consists of $\log_2N$ columns, each with $2N$ incident edges. A BP decoder traverses the graph column-by-column and each time $2N$ soft messages at the corresponding edges are updated. If the graph is traversed in a round-trip manner, a single-column decoder takes $2\log_2N-1$ clock cycles to complete an iteration.
A double-column architecture was proposed in \cite{Park2014} where the operations of two adjacent columns are merged in one clock cycle, effectively reducing in half the latency per iteration. Although the critical path was shown to increase by 14\% compared to that of the single-column architecture, the decoding throughput was improved by more than 40\%.

The graph can also be traversed uni-directionally, e.g., only activating columns from right to left. Under this schedule, data dependency is relaxed. Taking Fig.~\ref{fig:sc-graph} for example, the messages updated by the left-most column $c_0$ of the current iteration are not used by the right-most column $c_2$ of the next iteration. Hence, the operations of these two columns can be simultaneously executed and one clock cycle is saved.
More generally, an iteration-level overlapping schedule was proposed in \cite{Yuan2014c}. By increasing the hardware complexity to compute $\frac{1}{2}\log_2N$ columns simultaneously, the latency of a uni-directional BP decoder is reduced from $J\log_2N$ clock cycles to $2J+\log_2N-2$, where $J$ denotes the number of iterations.

\subsection{Low-Complexity BP Decoders}
\label{subsec:LC_BPD}
In \cite{Fayyaz2014}, a low-complexity variant of the BP decoding algorithm called soft cancellation (SCAN) was proposed. Under that algorithm, soft messages propagate according to the schedule of SC decoding, resulting in an increased latency in $\mathcal{O}(N)$ clock cycles.
However, the message propagation of SCAN was shown to be efficient and its overall complexity at low SNR regime was reduced by an order of magnitude compared to that of regular BP decoding.

Similar to iterative decoders for LDPC or turbo codes, early stopping schemes can be used in BP decoders for polar codes. 
In \cite{Yuan2014d}, threshold detections on messages are made after each iteration. When they satisfy certain criteria, decoding is stopped immediately. Furthermore, messages related to  sub-graphs are checked in \cite{Mohsin2015} such that the operations in those sub-graphs can be stopped earlier. 
As a result, with the method of \cite{Yuan2014d}, the average decoding complexity is reduced by around 30\% with negligible performance degradation. The method in \cite{Mohsin2015} further reduces the average complexity by 40\%.

\subsection{Memory-Efficient BP Decoders}
\label{subsec:ME_BPD}
It was observed in \cite{Park2014} that, if a single-column decoder follows a round-trip schedule, $N$ messages instead of $2N$ need to be updated each time. Thus, the memory requirement is reduced in half. 
In \cite{Sha2015}, two adjacent columns are combined into one so that intermediate messages need not to be stored. However, the corresponding message updating rules have to be modified accordingly. As a result, the overall memory usage is significantly reduced, at the cost of some combinational logic overhead.

\subsection{Implementation Results}
\label{subsec:IR_BPD}
Table~\ref{tab:impl:bpd} summarizes the ASIC implementation results of different BP decoders. It can be seen that, due to their parallel nature, the state of the art is already capable of achieving throughputs of multiple Gbps. However, even at high SNR, the average iteration number is high resulting in a greater computational complexity than its SC-based counterparts.

\begin{table}[t]
  \centering
  \caption{ASIC implementation results of BP decoders for a $(1024,512)$ polar code.}
  \vspace{-2.5pt}
  \begin{tabular}{lccc}
    \toprule
    Implementation & \cite{Park2014} & \cite{Yuan2014d} & \cite{Sha2015}\\
    \midrule
 Architecture & double-col. & overlapped & col.-combined\\
 Schedule & round-trip & uni-direction & round-trip\\
 Technology & 65 nm & 45 nm &  45 nm\\
 Area (mm$^2$) & 1.476 & N/A & 0.747\\
 Supply (V) & 1.0 & 1.1 & N/A\\
 $f$ (MHz)  & 300 & 500 & 197 \\
 Max. iter. \# & 15 & 40 & 15\\
 T/P (Gbps)  & 2.05 & 2.9 & 1.683\\
 Avg. iter. \#@SNR & 6.57@4.0 dB & 23.0@3.5 dB & N/A\\
 T/P (Gbps)@SNR  &4.68@4.0 dB & 4.5@3.5 dB & N/A\\
    \bottomrule
  \end{tabular}
  \label{tab:impl:bpd}
\end{table}


\section{Successive Cancellation List Decoders}
\label{sec:list}
In successive cancellation list (SCL) decoding, instead of decoding a single codeword, a list of $L$ tentative codewords (commonly called decoding \emph{paths}) is decoded simultaneously and the final codeword can be selected with the help of a CRC~\cite{Tal2015}. The list of $L$ paths can be processed in parallel to a large extent using up to $L$ SC decoders. In the simplest form of SCL decoding, the SC decoders only interact when a leaf node of the decoder tree that corresponds to an information bit is activated. When this happens, the SCL algorithm has to select the $L$ most likely paths out of $2L$ possible paths, and continues SC decoding with the surviving paths.

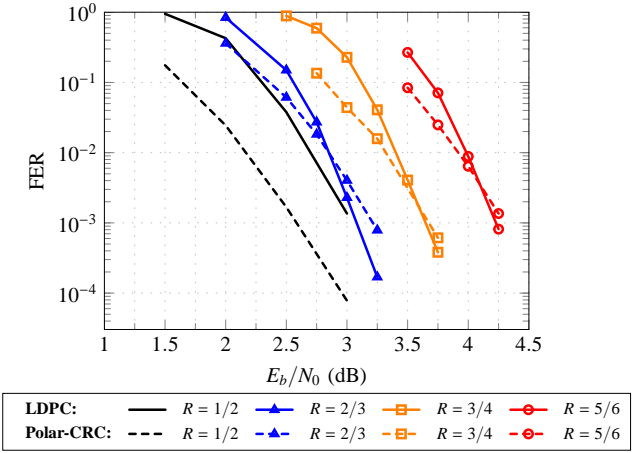
\begin{figure}[t]
  \centering
  \scalebox{0.8}{\begin{tikzpicture}
  \begin{semilogyaxis}
    [
    width=3.4in,
    height=2.7in,
    line cap=round,
    every axis y label/.style={at={(ticklabel cs:0.5)},rotate=90,anchor=near ticklabel},
    every axis x label/.style={at={(ticklabel cs:0.5)},anchor=near ticklabel},
    xmin=1,
    xmax=4.5,
    ymax=1,
    xlabel={$E_b/N_0$ (dB)},
    minor x tick num={1},
    ylabel={FER},
    legend columns = 5,
    legend style={
      at={(0.5, -0.2)},
      anchor={north},
      cells={anchor=west},
      column sep= 2mm,
      font=\footnotesize,
    },
    grid=major,
    xminorgrids=true,
    yminorgrids=false,
    grid style=loosely dotted,
    ]
    
    \addlegendimage{empty legend}
    \addlegendentry{\textbf{LDPC:}}

    \addplot[color=black,very thick] table[x=ebn0_db,y=FER] {data/wifi5.dat};
    \addlegendentry{$R = 1/2$}
    \addplot[color=blue,very thick, mark=triangle, mark options={solid}] table[x=ebn0_db,y=FER] {data/wifi67.dat};
    \addlegendentry{$R = 2/3$}
    \addplot[color=orange,very thick, mark=square, mark options={solid}] table[x=ebn0_db,y=FER] {data/wifi75.dat};
    \addlegendentry{$R = 3/4$}
    \addplot[color=red,very thick, mark=o, mark options={solid}] table[x=ebn0_db,y=FER] {data/wifi83.dat};
    \addlegendentry{$R = 5/6$}

    \addlegendimage{empty legend}
    \addlegendentry{\textbf{Polar-CRC:}}

    \addplot[color=black,very thick, dashed] table[x=ebn0_db,y=FER] {data/crc8-1k-l2.dat};
    \addlegendentry{$R = 1/2$}

    \addplot[color=blue,very thick,dashed, mark=triangle, mark options={solid}] table[x=ebn0_db,y=FER] {data/crc8-1k-l2-66.dat};
    \addlegendentry{$R = 2/3$}

    \addplot[color=orange,very thick,dashed, mark=square, mark options={solid}] table[x=ebn0_db,y=FER] {data/crc8-1k-l2-75.dat};
    \addlegendentry{$R = 3/4$}

    \addplot[color=red,very thick,dashed, mark=o, mark options={solid}] table[x=ebn0_db,y=FER] {data/crc8-1k-l2-83.dat};
    \addlegendentry{$R = 5/6$}

  \end{semilogyaxis}

\end{tikzpicture}}
  \vspace{-2.5pt}
  \caption{Frame error rate of polar codes of length $N = 1,024$ under SCL decoding with $L=2$ and an $8$-bit CRC compared with the LDPC codes of the IEEE 802.11n standard of length $N=1,944$ under offset min-sum decoding with a flooding schedule and a maximum of $10$ iterations. All simulations were performed using BPSK modulation over an AWGN channel.}
  \label{fig:wifi-perf}
\end{figure}

In Fig.~\ref{fig:wifi-perf}, we observe that polar codes under SCL decoding can achieve similar performance to the LDPC codes used in the IEEE 802.11n standard.

\subsection{Exact LLR-Based SCL Decoding}
The original description of SCL decoding was made using channel \emph{likelihoods}~\cite{Tal2015}. While such a high-level description is valid, these likelihoods can become very small during the decoding process, resulting in numerical precision problems and inefficient hardware implementations. The first hardware implementations of SCL decoding used log-likelihoods to partially overcome these problems~\cite{Balatsoukas2014,Lin2014a}. An equivalent, but much more efficient, description of SCL decoding in terms of LLRs and an LLR-based \emph{path metric} was presented in~\cite{Balatsoukas2015,Yuan2014b}. LLR-based SCL decoding leads to the most efficient exact\footnote{The implementation is ``exact'' up to quantization loss and min-sum approximation loss, which are common losses to all hardware implementations.} hardware implementation of SCL decoding~\cite{Balatsoukas2015}.

\subsection{Path Metric Sorting}
A computationally challenging step of SCL decoding is that of path metric sorting, where the $L$ best path metrics have to be selected among $2L$ possible metrics. The properties of the LLR-based path metric were exploited in~\cite{Balatsoukas2015b} in order to significantly reduce the complexity and the critical path of the metric sorting blocks, while still performing exact sorting.

In a different approach, path metrics were approximately sorted with double thresholding method proposed in~\cite{Fan2015}. It compares $2L$ path metrics with two thresholds, $AT$ and $RT$. A path survives if its metric is smaller than $AT$. After it, the paths with metrics in between $AT$ and $RT$ are randomly selected to fill up the list of $L$ surviving paths. This method significantly reduces the critical path of metric sorting, especially for a large $L$. 
The problem of path metric sorting is even more pronounced in decoders that employ multi-bit decision, such as \cite{Yuan2014,Lin2014}, since in this case the $L$ best metrics out of up to $2^bL$ candidate metrics have to be selected, where $b$ is the number of bits that are decoded simultaneously. To this end, an approximate two-stage sorting unit was proposed in~\cite{Xiong2015}, where the best $q$ out of $2^b$ successor paths are first selected for each path, and then the best $L$ paths are selected among the $qL$ paths resulting from the first step. This approach reduces the sorting complexity at the cost of a small performance degradation.

\subsection{Approximate Tree Pruning in LLR-Based SCL Decoding}
Since SCL decoding heavily relies on SC decoding for the computation of the path metrics, one may expect that the pruning techniques described in Section~\ref{sec:sc-based} should be applicable to SCL decoding as well. Unfortunately, in order to update the LLR-based path metric of~\cite{Balatsoukas2015}, all the LLRs produced by the leaves of the decoder tree are required and these node computations cannot be pruned if one wants to implement exact SCL decoding. Nevertheless, \cite{Lin2014} describes an approximate SCL decoding algorithm which is based on tree pruning that simply ignores some of the path metric updates corresponding to computation tree leaves. The resulting performance degradation is small and an outline of a hardware architecture is presented which can achieve a throughput of slightly over $1$~Gbps for a polar code of length $N = 8192$ and rate $R = 0.5$ using SCL decoding with $L = 4$.

\subsection{Implementation Results}
Table~\ref{tab:sclresults} summarizes the implementation results of the aforementioned state-of-the-art SCL decoder architectures. While SCL decoders have an error-correction performance that is close to (or even better than) LDPC codes, we observe that the implementation of multi-Gbps SCL decoders remains a challenging problem.

\begin{table}[t]
	\centering
	\caption{ASIC implentation results for various SCL decoders.}\label{tab:sclresults}
  \vspace{-2.5pt}
	\begin{tabular}{lcccc}
	\toprule
	Implementation    & \cite{Balatsoukas2015}	& \cite{Xiong2015}	& \cite{Fan2015}	& \cite{Lin2014} \\
	\hline
	Code Length   		&	1,024										& 1,024							& 1,024						& 8,192 \\
	Rate 							&	0.5											& 0.5 							& 0.5							& 0.5 \\
	List Size 				&	4												& 4									& 16							& 4 \\
	Algorithm	    		& Exact										& Approx. 					& Approx.					& Approx. \\
	Technology				&	90 nm										& 90 nm 						& 90 nm						& 90 nm \\
	Area (mm$^2$)			&	1.78										&	1.21							& 7.46						& 2.45 \\
  Supply (V)        & 1.0                     & N/A               & 1.2             & N/A \\
  $f$ (MHz)         & 794                     & 500               & 641             & 400 \\
	T/P (Mbps)	      &	307											& 313								& 220							& 1,052 \\
	\bottomrule
	\end{tabular}
\end{table}


\section{Conclusion}
\label{sec:conclusion}
In this paper, we reviewed the state of the art in polar decoders implementing the successive-cancellation, list, and belief propagation decoding algorithms. The advantages of the different algorithms were illustrated. It was shown that the many decoding algorithms were developed and implemented to address various application requirements: from error-correction performance rivaling that of LDPC codes to very high throughput or low-complexity decoders. 

\bibliographystyle{IEEEtran}
\bibliography{IEEEabrv,polar-overview}

\end{document}